\documentclass{gNST2e}

\usepackage{epstopdf}
\usepackage{subfigure}

\theoremstyle{plain}
\newtheorem{proposition}{Proposition}
\newtheorem{result}{Result}

\newtheorem{remark}{Remark}

\theoremstyle{definition}

\usepackage{amssymb}
\usepackage{amsmath}
\usepackage{bm}
\usepackage{geometry}

\usepackage{hyperref}
\hypersetup{
    colorlinks  = true,     citecolor = blue
}

\usepackage{graphicx} 
\graphicspath{{figures/}}
\usepackage{enumitem}

\usepackage{soul}

\begin{document}



\title{Profile likelihood ratio tests for parameter inferences in \\ generalized single-index models}

\author{
\name{Nanxi Zhang\textsuperscript{a} and Alan Huang$^{\ast, a}$\thanks{$^\ast$Corresponding author. Email: alan.huang@uq.edu.au}}
\affil{\textsuperscript{a}School of Mathematics and Physics, The University of Queensland, Qld 4072, Australia}
}

\maketitle

\begin{abstract}
A profile likelihood ratio test is proposed for inferences on the index coefficients in generalized single-index models. Key features include its simplicity in implementation, invariance against parametrization, and exhibiting substantially less bias than standard Wald-tests in finite-sample settings. Moreover, the R routine to carry out the profile likelihood ratio test is demonstrated to be over two orders of magnitude faster than the recently proposed generalized likelihood ratio test based on kernel regression. The advantages of the method are demonstrated on various simulations and a data analysis example.
\end{abstract}

\begin{keywords}
Generalized single-index models; regression splines; profile likelihood ratio test; parameter inference
\end{keywords}

\begin{classcode}
62F03; 62J12 
\end{classcode}

\section{Introduction}
\label{s1}
Single-index models \citep[SIMs, e.g.,][]{Hard:Hall:Ichi:optm:1993, Ichimura:1993} are extremely useful tools for analysing complex multivariate data in a parsimonious yet flexible way. SIMs make the simplifying assumption that the conditional distribution of a response $Y$ given a set of covariates $\bm{x} \in \mathbb{R}^d$ depends only on the single index $\bm{x}^T \bm{\beta}$ for some vector of parameters $\bm{\beta} \in \mathbb{R}^d$, with the functional form of this dependency left unspecified. It is this balance of model parsimony, interpretability and flexibility that has seen SIMs find a wide range of applications in a variety of fields, such as in econometrics \citep[e.g.,][]{Horowitz:2009, Hu:Shiu:Wout:iden:2015} and survival analysis \citep[e.g.,][]{Hard:Hall:Ichi:optm:1993, SK:2014}.

In this note we consider generalized single-index models (GSIMs) which assume that the conditional distribution of a response $Y$ given a set of covariates $\bm{x}$ has the form
\begin{equation}
\label{eq:pdf}
Y \mid \bm{x} \sim f(y | \bm{x}; \bm{\beta}) = \exp \left\{\frac{y g(\bm{x}^T \bm{\beta}) - b(g(\bm{x}^T \bm{\beta}))}{\varphi} + c(y; \varphi) \right\},
\end{equation}
where the functions $b(\cdot)$ and $c(\cdot)$ are of known forms, $g: \mathbb{R} \rightarrow \mathbb{R}$ is a smooth but otherwise unspecified function, $\bm{\beta}$ is a vector of coefficients, and $\varphi$ is a dispersion parameter. This framework covers normal, Poisson, binomial and gamma responses, amongst others. Model (\ref{eq:pdf}) implies $E(Y|\bm{x}) = b'(g(\bm{x}^T \bm{\beta}))$, where $b'(\cdot)$ is the canonical inverse-link function. As with classical generalized linear models (GLMs), the use of the canonical link generally leads to sensible conditional mean functions for any function $g$. For example, $b'(\cdot) = \exp(\cdot)$ for Poisson regression, ensuring non-negativity of the conditional mean for any value of the function $g$. For binomial responses, the canonical link $b'(\cdot) = \exp(\cdot)/(1 + \exp(\cdot))$ ensures that the conditional mean is between 0 and 1 for any value of the function $g$.

The computational aspects of fitting GSIMs have been widely discussed in the literature. The terminology ``bundled parameter" was first used in \citet{Huang:1997} to describe $(\bm{\beta}, g(\cdot; \bm{\beta}))$, where the finite-dimensional index coefficients $\bm{\beta}$ of interest and the infinite-dimensional nuisance parameter $g(\cdot)$ are bundled together. Various methods are available for estimating $(\bm{\beta}, g(\cdot))$ simultaneously, for example, kernel smoothing \citep{Weisberg:1994}, average derivative estimation \citep{Hardle:1989}, sliced inverse regression \citep{YinCook2005}, local linear methods \citep{Carr:Fan:Gijb:Wand:gene:1997}, and penalized splines \citep{Yu:Wu:Zhang:pen:2017}. Implicit to any fitting method is the selection of a smoothing parameter, which is used to prevent overfitting and to regularize the underlying computational problem. The smoothing parameter is usually chosen via generalized cross-validation \citep[e.g.,][]{Yu:Wu:Zhang:pen:2017}, or set to some ``optimal" value relative to a working model \citep[e.g.,][]{Zhan:Huan:Lv:stat:2010}. All the above fitting methods have their own merits, and their corresponding estimators have competing attractive properties -- see individual references for details on each method.

While most of the existing literature on GSIMs focus on model estimation and smoothing parameter selection, inferences on the index parameters $\bm{\beta}$ are less explored. This paper is mainly concerned with this latter problem. To this end, we propose a profile likelihood ratio test (PLRT) for testing the index parameters that is simple on both a conceptual and computational level. Conceptually, to test between two nested models we simply fit both models and compute a likelihood ratio statistic between the two models. This is then compared to the usual asymptotic $\chi^2$ distribution, or an $F$ distribution for a finite-sample adjustment. Computationally, the PLRT involves no more than adding a few lines of code to existing software for fitting GSIMs. For this paper, we recommend the \texttt{gam} (generalized additive models) function in the state-of-the-art R package \texttt{mgcv} \citep{Wood:2016}, although other software can be modified in a similar way to carry out the proposed PLRT procedure. The R code to carry out these computations is particularly simple, and can be downloaded from the Online Supplement.

A reviewer pointed out that our model estimation procedure is similar to that in \citet{Yu:Wu:Zhang:pen:2017}, where generalized partially linear single-index models 
are investigated. Indeed, both papers carry out model fitting via penalized splines. However, \citet{Yu:Wu:Zhang:pen:2017} focus on model fitting and parameter estimation, while the main focus here is on parameter inferences. Although \citet{Yu:Wu:Zhang:pen:2017} establish large-sample properties for their estimator and propose a sandwich formula for estimating the asymptotic variances for Wald-based inferences, the practical performance of their approach is not examined in their numerical or data analysis examples. The proposed PLRT also enjoys some unique advantages over Wald-based inferences, which we now highlight.

In addition to being conceptually and computationally simple, the proposed PLRT method is also invariant to model parametrization. A well-known property of SIMs and GSIMs is that they are not generally identifiable. Two popular sets of identifiability constraints on $\bm{\beta}$ are:
\begin{enumerate}
\item $\bm{\beta}$ contains no intercept term, $\beta_1 > 0$, and $\| \bm{\beta} \|^2 = 1$ \citep[e.g.,][]{Yu:2002, Hard:Mull:Sper:Werw:nonp:2004, Zhan:Huan:Lv:stat:2010, Cui:Hard:Zhu:efm:2011},
\item $\bm{\beta}$ contains no intercept term, and $\beta_1 = 1$  \citep[e.g.,][]{Hard:Mull:Sper:Werw:nonp:2004}.
\end{enumerate}
Although the fitted model is the same under any set of identifiability constraints, inferences based on Wald-tests are generally not invariant to parametrization. For example, in either parametrization above it is not possible to test $\beta_1 = 0$, that is, if covariate $x_1$ has no overall effect on $Y$. In parametrization 2, it is also not possible to compute standard errors for the estimated coefficient for $x_1$ as $\beta_1$ is always set to 1. Instead, one needs to relabel the covariates so that $x_1$ no longer corresponds to the first coefficient, in order to carry out inferences on the effect of covariate $x_1$. In contrast, the proposed PLRT is invariant to parametrization as it exploits the fact that the fitted model, and subsequently the maximized likelihood, is the same regardless of parametrization. Thus, we can simply fit the model with and without the covariate $x_1$ and compare the maximal log-likelihoods achieved, regardless of which identifiability constraint is used.

The PLRT approach also exhibits substantially less bias than the usual Wald tests in all our simulation settings (see Section \ref{s3}). We suspect that this is because the proposed PLRT bypasses explicit estimation of the variance matrix which is at the crux of Wald-based inferences. This variance matrix is typically estimated by plugging in the estimated $\hat{\bm{\beta}}$ and $\hat g$ into the expression for the asymptotic variance \citep[e.g.,][Section 4]{Yu:Wu:Zhang:pen:2017}. The estimation of $g$ is generally very noisy, which leads to inaccurate variance estimation and subsequently biased Wald statistics. The level of bias can be severe, as demonstrated in our simulations in Section \ref{s3}. We stress that we use only the default automated smoothing parameter selection from the \texttt{gam} function when implementing the proposed PLRT. In particular, we never ``hand-pick" a smoothing parameter value to make our method look superior to competing methods in any of our simulations or data analysis example. 

\section{Related methods}
There are two closely related approaches for inferences on index coefficients using likelihood-type functions. These are the generalized likelihood ratio test \citep[GLRT;][]{Zhan:Huan:Lv:stat:2010} and the conditional quasi-likelihood ratio test \citep[QLRT;][]{Cui:Hard:Zhu:efm:2011}. We compare and contrast these methods here.

\subsection{Generalized likelihood ratio test} 
The GLRT approach of \citet{Zhan:Huan:Lv:stat:2010} employs local-linear estimation for the function $g$ in the special case of additive errors with constant variance. This is done via the following three steps:
\begin{enumerate}[leftmargin=*]
\item For each $z$ and $\boldsymbol{\beta}$, minimize $\sum_{i=1}^n  [Y_i - a - b(\boldsymbol{x}_i^T \boldsymbol{\beta} - z)]^2 K_h(\boldsymbol{x}_i^T \boldsymbol{\beta} - z)$ in $a$ and $b$, where $K_h$ is some kernel function with bandwidth $h$, giving local estimates of the intercept $\hat a = \hat a(z; \boldsymbol{\beta}, h)$ and slope $\hat b = \hat b(z; \boldsymbol{\beta}, h)$.
\item Minimize the residual sum of squares $\sum_{i=1}^n  [Y_i - \hat a(\boldsymbol{x}_i^T\boldsymbol{\beta}; \boldsymbol{\beta}, h)]^2$ in $\boldsymbol{\beta}$, subject to identifiability constraint $\boldsymbol{\beta}^T \boldsymbol{\beta} = 1$. This gives the estimate $\hat{\boldsymbol{\beta}}$.
\item Estimate $g$ by $\hat g(\cdot \ ; h) = \hat a(\cdot \ ; \hat{\boldsymbol{\beta}}, h)$.
\end{enumerate}

To test the null hypothesis $H_0: \boldsymbol{\beta}_{\{l\}} = 0 $ against the alternative $H_1: \mbox{ not all } \boldsymbol{\beta}_{\{l\}} = 0$, where $l \subset \{1,2,,\ldots, d\}$ is some subset of indices, the GLRT proceeds by carrying out the above three steps under both $H_0$ and $H_1$, and computing the log ratio of the residual sum of squares. This can then be shown to follow
a scaled asymptotic $\chi^2$ distribution, with the scaling factor and degrees of freedom depending on the kernel function $K$, the bandwidth $h$, and the support of the estimated linear predictor $\boldsymbol{x}_i^T \hat{ \boldsymbol{\beta}}$ under both the null and alternative hypotheses.

To carry out the GLRT in practice, \citet{Zhan:Huan:Lv:stat:2010} suggest two tweaks to the theory. First, instead of directly using the asymptotic $\chi^2$ result, the authors recommend bootstrap resampling to estimate the quantiles of the null distribution. This is because the null distribution depends on the estimated support from both the null and alternative fitted models. In this sense, the GLRT exhibits a non-standard type of Wilks phenonenom. Second, the recommended bandwidth for hypothesis testing is different to the optimal bandwidth for fitting the model. More precisely, if $\hat{h}_\text{opt}$ is the estimated optimal bandwidth for fitting the model, then the corresponding optimal bandwidth for hypothesis testing was found to be $\hat{h}_\text{opt} \times n^{-1/20}$ numerically. This treats model fitting and model inferences on slightly different footings. These two tweaks were employed throughout the simulation studies and data analysis example in \citet{Zhan:Huan:Lv:stat:2010}. 

The main advantage of the proposed PLRT framework over the GLRT is that it is simpler to implement in practice. In particular, the asymptotic distribution for calibrating the test does not depend on a chosen kernel function, a chosen bandwidth, nor the support of the fitted linear predictors. It also does not require bootstrap approximations for the null distribution, nor tweaking of the bandwidth -- in fact, we use only the default automated smoothing parameter selection from the \texttt{gam} function from the \texttt{mgcv} package \citep{Wood:2016}. That is, we simply fit the model under both the null and alternative hypotheses using the default automated smoothing parameter selection, and directly compare the likelihood ratio statistic to an asymptotic $\chi^2$ distribution, with degrees of freedom depending only on the number of constraints imposed by the null hypothesis. Thus, PLRT is much more computationally efficient than GLRT. Indeed, our numerical examples in Section \ref{s3} demonstrate that the proposed PLRT is over two magnitudes of order faster to carry out than the GLRT approach. 

The GLRT approach is also inappropriate for data with non-constant variance, which is typical of count, binomial and time-to-event responses. However, such responses pose no problems for the proposed PLRT approach as it is based on the generalized linear model framework \citep{McCullagh:1989}.

Finally, the code for implementing the GLRT is not readily available, even after contacting the authors. To this end, we emulated the approach in R using the \texttt{npindex} function from the \texttt{np} package \citep{RH:2016}, and we used this implementation for our simulation studies in Section \ref{s3}. Our replica code is provided in the Online Supplement.

\subsection{Conditional quasi-likelihood ratio test}
For handling data with non-constant variance, \citet{Cui:Hard:Zhu:efm:2011} replace the sum of squares criterion in the above three steps from \citet{Zhan:Huan:Lv:stat:2010} with a quasi-likelihood criterion specified via mean-variance relationship. A conditional quasi-likelihood ratio test (QLRT) can then be constructed for inferences on the index coefficients $\bm{\beta}$.


More precisely, to test the null hypothesis $H_0: \boldsymbol{\beta}_{\{l\}} = 0 $ against the alternative $H_1: \mbox{not all } \bm{\beta}_{\{l\}} = 0$, where $l \subset \{1,2,,\ldots, d\}$ is some subset of indices, the QLRT first fits a local linear quasi-likelihood model under $H_1$. Then, conditional on the fitted smooth function $\hat g$ obtained under $H_1$, a second quasi-likelihood model under $H_0$ is fitted.
A quasi-likelihood ratio statistic between the two models fits is computed, which can then be compared to an asymptotic $\chi^2$ distribution with degrees of freedom given by the number of constraints imposed by $H_0$. This approach is conditional because the fitted smooth function $\hat g$ under the alternative hypothesis is treated as fixed under the null hypothesis and also in the subsequent quasi-likelihood ratio statistic. In contrast, the proposed PLRT approach is an unconditional test as it refits the smooth function $g$ and the coefficients $\bm{\beta}$ under both the null and alternative hypotheses.

In practice, the QLRT also differs from the proposed approach as it requires selection of an additional adjustment factor to enhance the stability and accuracy of corresponding algorithm. \citet{Cui:Hard:Zhu:efm:2011} suggest numerically searching for the ``optimal" value of this adjustment factor over some interval determined by the dimension of the problem, with the criterion for being ``optimal" defined relative to some assumed working model. In contrast, the proposed PLRT does not require any additional stability parameter.  

Moreover, while the asymptotic theory for the QLRT is valid for any well-behaving bandwidth selection method, such as cross-validation, the actual bandwidth selection method used throughout the simulation studies in \citet{Cui:Hard:Zhu:efm:2011} seems to be fine-tuned using knowledge of the true underlying function $g$. For real data analysis problems where the true curve is unknown, the authors recommend ``trying a number of smoothing parameters that smooth the data and picking the one that seems most reasonable". This approach can be subjective and ambiguous. In contrast, the PLRT approach we examine here is implemented in the same automated way in all of our simulations and data analysis examples. In particular, we never fine-tune the smoothing parameter using knowledge of the true curve in any of our numerical studies. The R code to implement the PLRT is also particularly simple.

\vspace{-2.2mm}

\section{Model and main results}
\label{s2}
\subsection{Model and estimation}
A wide range of nonparametric estimation approaches exist for fitting generalized single-index models (\ref{eq:pdf}) to data, including kernel and local polynomial regression \citep{Cui:Hard:Zhu:efm:2011} and sliced inverse regression \citep{YinCook2005}. In this paper, we consider penalized regression splines for both model fitting and parameter inferences. We find penalized splines particularly simple to work with on both a theoretical and practical level. 

More precisely, the smooth function $g(\cdot)$ is approximated by a series expansion,
$
g(\cdot) = \bm{\delta}^T B(\cdot) \ ,
$
where $\bm{\delta}$ is a vector of spline coefficients, and $B(\cdot)$ is a set of basis functions. Various types of basis functions can be used here, with the two most popular choices being cubic regressions splines \citep[Section 4.1.2]{Wood:2006} and truncated P-splines \citep{Yu:2002, Yu:Wu:Zhang:pen:2017}. The theory and methodology in this paper are valid for both of these approaches.

For parameter identifiability in model estimation, we use the first set of constraints from Section \ref{s1}. That is, the parameter space of $\bm{\beta}$ is $\{\bm{\beta} = (\beta_1, \dots, \beta_d)^T: \| \bm{\beta} \|^2 = 1, \beta_1 > 0, \bm{\beta} \in \mathbb{R}^d \}$, where $\| \cdot \|$ denotes the Euclidean norm. The parameter $\bm{\beta}$ is on the boundary of a unit ball, which violates the usual regularity conditions needed to establish asymptotic properties of subsequent estimators \citep[Section 2]{Cui:Hard:Zhu:efm:2011}. By introducing a $(d-1)$-dimensional parameter $\bm{\phi} = (\phi_1, \ldots, \phi_{d-1})^T$, we can parametrize $\bm{\beta}$ through $\bm{\beta}(\bm{\phi}) = (\sqrt{1-\| \bm{\phi} \|^2}, \phi_1, \ldots, \phi_{d-1})^T$, where $\bm{\phi}$ satisfies the constraint $\| \bm{\phi} \| \le 1$. If the true value $\bm{\phi}_*$ is such that $\|\bm{\phi}_*\| <1$, then standard regularity conditions hold. 

\begin{remark}
Identifiability constraints are only needed for model estimation. The fitted model and, subsequently, the likelihood value achieved are the same regardless of which set of identifiability constraints is used. Thus, parameter inferences based on the likelihood are invariant to parametrization. It is this key property that we exploit in Section \ref{se:LRT} of this paper.
\end{remark}

A penalized likelihood estimator of $\bm{\theta} = (\bm{\phi}^T, \bm{\delta}^T)^T$ can then be obtained by maximizing the penalized log-likelihood function,
\begin{eqnarray}\label{eq:obj}
  \ell_{n \lambda}(\bm{\theta}) = \ell_n(\bm{\theta}) - \frac{n}{2}\lambda_n \bm{\delta}^T D \bm{\delta} \ ,
\end{eqnarray}
where $\ell_n(\bm{\theta}) = \sum_{i=1}^n [y_i \bm{\delta}^T B(\bm{x}_i^T \bm{\beta}(\bm{\phi})) - b(\bm{\delta}^T B(\bm{x}_i^T \bm{\beta}(\bm{\phi})))]$ is the unscaled log-likelihood, $\lambda_n \geq 0$ is a smoothing parameter, and $D$ is a positive semi-definite symmetric matrix satisfying
$
\bm{\delta}^T D \bm{\delta} = \int [g''(z)]^2 dz \ .
$
This penalizes the curvature of $g$ to avoid the overfitting of regression curve. A smaller value of $\lambda$ results in a more wiggly fitted function $\hat{g}$ that may capture local fluctuations, while a larger value of $\lambda$ leads to an increasingly linear estimation of function $g$. 

Finally, the dispersion parameter $\varphi$ can be estimated from the Pearson residuals using the method-of-moments estimator,
$
\hat \varphi = (n-k)^{-1}\sum_{i = 1}^n (Y_i - \hat{\mu}_i)^2/\hat v_i
$,
where $k$ is the degree of the freedom of the fitted model, $\hat{\mu}_i$ are the estimated means, and $\hat v_i = b''(\hat{\boldsymbol{\delta}} B(\boldsymbol{x}_i^T \boldsymbol{\beta}(\hat{ \boldsymbol{\phi}})))$ are the estimated (unscaled) variances.

\subsection{Large sample properties}
We follow the fixed-knot asymptotics of \citet{Yu:2002} and assume that the true underlying function $g$ is itself a spline function. For functions $g$ that are not spline functions, the asymptotic bias can be offset by increasing the number of knots. However, as \citet[Section 3]{Yu:2002} argue, the variability in the choice of smoothing parameter in practice is typically larger than this asymptotic bias and so fixed-knot asymptotics are a reasonable approximation for practical purposes. The assumptions we impose on $\bm{\theta} = (\bm{\phi}^T, \bm{\delta}^T)^T$ and the corresponding parametrized space $\Theta$ are specified in the Appendix. 
Results 1 and 2 below follow from \citet{Yu:2002}.

\begin{result}[Consistency] Under Assumptions A1--A3 in the Appendix, if the smoothing parameter $\lambda_n = o(1)$ then there exists a local maximizer $\hat{\bm{\theta}}$ of (\ref{eq:obj}) such that $\|\hat{\bm{\theta}} - \bm{\theta} \| = O_p(n^{-1/2}+\lambda_n)$. In particular, $\hat{\bm{\theta}} \rightarrow \bm{\theta}$ in probability.
\end{result}

\begin{result}[Asymptotic normality] Under Assumptions A1--A3  in the Appendix, if the smoothing parameter $\lambda_n = o(n^{-1/2})$ then a sequence of constrained penalized estimators $\hat{\bm{\theta}} = (\hat{\bm{\phi}}^T, \hat{\bm{\delta}}^T)^T$ exists, is consistent, and is asymptotically normally distributed. That is,
$
\sqrt{n}(\hat{\bm{\theta}} - \bm{\theta}_*) \rightarrow N\left(0, I(\bm{\theta}_*)^{-1}\right)
$
in distribution, where $I(\theta_*)$ is the Fisher information matrix defined in the Appendix. Moreover, we have
\begin{equation}
\label{eq:asymnorm}
\sqrt{n}\left(
  \begin{array}{c}
    \hat{\bm{\beta}} - \bm{\beta} \\
    \hat{\bm{\delta}} - \bm{\delta} \\
  \end{array}
\right)
 \rightarrow
N\left(0, J(\bm{\theta})I(\bm{\theta})^{-1}J(\bm{\theta})^T\right)
\end{equation}
in distribution, where $J$ is the Jacobian matrix for transforming back from $\bm{\theta} = (\bm{\phi}^T,\bm{\delta}^T)^T$ to $(\bm{\beta}^T,\bm{\delta}^T)^T$.
\end{result}

Result 2 is often used to motivate Wald statistics for inferences on the regression parameters $\bm{\beta}$, with the asymptotic variance in (\ref{eq:asymnorm}) estimated using a plug-in estimator by substituting the fitted $\hat{\bm{\delta}}$ and $\hat{\bm{\beta}}$ in for $\bm{\delta}$ and $\bm{\beta}$. However, Wald tests using a plug-in estimator of variance can be very biased in practice, as demonstrated in our simulations in Section \ref{s3}. We suspect that this is due to the fact that $\hat g$ can still exhibit a lot of local fluctuations even with large sample sizes. Another drawback is that the Wald-test is not invariant to the choice of identifiability constraints. As mentioned in Section \ref{s1}, it is not possible to test if $\beta_1 = 0$, that is, if covariate $x_1$ has no overall effect on $Y$, without first reparametrizing the model so that $x_1$ is no longer the first covariate.

\subsection{Profile likelihood ratio test}
\label{se:LRT}
To overcome the drawbacks of the Wald-test, we propose an alternative approach for inferences on $\bm{\beta}$ that does not require explicit estimation of the variance, is easy to implement computationally, and is invariant to identifiability constraints. The method is based on the profile loglikelihood function for $\bm{\beta}$, which is defined as
$$
pl(\bm{\beta}) =  \sum_{i=1}^n [y_i \hat{\bm{\delta}}_\beta^T B(\bm{x}_i^T \bm{\beta}) - b(\hat{\bm{\delta}}_\beta^T B(\bm{x}_i^T \bm{\beta}))] \ ,
$$
where $\hat{\bm{\delta}}_\beta$ is the maximizer of the penalized log-likelihood (\ref{eq:obj}) for fixed $\bm{\beta}$. A profile likelihood ratio test (PLRT) statistic can be then be constructed by comparing the profile likelihoods achieved under the null and alternative hypotheses. 

More precisely, suppose we are interested in testing the hypothesis $H_0: M\bm{\beta} = 0$ versus $H_1: M\bm{\beta} \neq 0$,
where $M$ is a $r \times d$ matrix with rank $r < d$ and $MM^T = I$. For example, if we are testing whether $x_1$ and $x_3$ have no overall joint effect on the response $Y$, then $r = 2$ and $M $ is
$$
M = \left(
  \begin{array}{cccccc}
    1 & 0 & 0 & 0 & \cdots & 0 \\
    0 & 0 & 1 & 0 & \cdots & 0
  \end{array}
\right).
$$
To carry out this test, we simply fit two models, one with the constraint $M \bm{\beta} = 0$ and one without, and evaluate the maximum profile likelihoods under the null and alternative hypotheses. The profile likelihood ratio statistic can then be shown to have usual $\chi^2$ asymptotic distributions. The proof of Proposition 1 is given in the Supplemental Materials.


\begin{proposition}[Profile likelihood ratio test] Suppose Assumptions A1--A3 in the Appendix hold and $\lambda_n = o(n^{-1/2})$. Then under the null hypothesis $H_0$,
$
2\left\{\sup_{H_1} pl(\bm{\beta}) - \sup_{H_0} pl(\bm{\beta}) \right\} \rightarrow \varphi \, \chi_r^2
$
in distribution as $n \rightarrow \infty$.
\label{prop:LRT}
\end{proposition}

In practice, $\varphi$ is typically unknown and we replace it with its estimate $\hat \varphi$. A finite-sample adjustment to the above test is to compare the profile likelihood ratio to an $rF_{r, \ n-df(H_1)}$ distribution instead, where $df(H_1)$ is the degrees of freedom of the fitted model obtained under alternative hypothesis. This is justified since $rF_{r, \ n - df(H_1)} = \chi_r^2 + o_P(1)$ for large $n$.

We can also use the PLRT to define equivalent standard errors for $\hat \beta_j$ via
\begin{equation}
\label{eq:se}
\mbox{se}_{eq}(\hat{\beta}_j) = \frac{ \sqrt{\hat{\varphi}} \, |\hat{\beta}_j|}{\sqrt{2\left\{\sup pl(\bm{\beta}) - \sup_{\beta_j = 0} pl(\bm{\beta}) \right\}}}
\end{equation}
where $\sup_{\beta_j = 0} pl(\bm{\beta})$ is the maximal log-likelihood achieved under the constraint $\beta_j = 0$. By construction, the $t$-statistic $|\hat \beta_j|/\mbox{se}_{eq}(\hat \beta_j)$ achieves the same significance as the PLRT for testing $\beta_j = 0$. A null value other than 0 can also be used to calculate the equivalent standard error, but in the absence of any additional knowledge about the true parameter value, the choice of 0 is a good default to use in practice.

\section{Simulation studies}
\label{s3}
To assess the practical performance of the proposed PLRT approach for inference on the index parameters $\bm{\beta}$, we looked at five sets of simulations covering continuous and binary responses, and monotonic, unimodal and sinusoidal means curves.
We employed cubic regression splines for the first two sets of simulations and truncated cubic splines for the other three, demonstrating that the methodology works well for either choice of basis functions. 
For monotonic or unimodal regressions, we follow the recommendation in \citet{Yu:2002} and set the default number of knots to $10$. For more complex regressions, the number of knots may be increased -- see \citet{Yu:2002} and \citet{Ruppert:2002} for more discussions on selecting the number of knots.

The practical performance of the proposed PLRT approach was compared to that of the standard Wald test, as well as that of the generalized likelihood ratio test (GLRT) of \citet{Zhan:Huan:Lv:stat:2010}. Interestingly, computer software for implementing the GLRT was not readily available, even after contacting the authors. For the purposes of this paper, we replicated the GRLT method ourselves in \texttt{R} using the \texttt{npindex} function from the \texttt{np} package \citep{RH:2016}. We employed local constant estimation using second order Epanechnikov kernels. All three methods were run on a Windows desktop with an i7-3770 CPU running at 3.40 GHz and 16.0 GB RAM. 


\subsection{Continuous responses with sinusoidal means}
To compare the performance of the GLRT, Wald and PLRT approaches for continuous data, we generated synthetic datasets using the sinusoidal model from \citet{Cui:Hard:Zhu:efm:2011}, 
$$Y_i \ |\ \bm{x}_i \sim N(\sin(a \bm{x}_i^T \bm{\beta}), \, \sigma^2) \ , \quad \mbox{ for } i=1,2,\ldots,n \ ,$$
with sample sizes $n=100$ and  $400$ covering moderately small to moderately large sample sizes. The true index parameters were set to $\bm{\beta} = (\beta_1,\beta_2, \beta_3, \ldots,\beta_{10})^T = (2,1,0,\ldots,0)^T / \sqrt{5}$. Each covariate in $\bm{x}_i$ were generated independently from a $N(2, 1)$ distribution and the error standard deviation $\sigma$ was set to $0.2$. Two different periodicities were considered, with $a = \pi/2$ corresponding to a unimodal mean function and $a = 3\pi/4$ corresponding to a mean function with one peak and one trough. A total of $N = 1000$ simulations were carried out for each setting.

For each simulated dataset, a Gaussian GSIM model was fit using either local linear estimation for the GLRT approach, or penalized cubic regression splines for the Wald and PLRT approaches. The bandwidth for the local linear approach was chosen via least-squares cross-validation method as implemented in the \texttt{np} package, while the smoothing parameter for penalized cubic splines was chosen via the default cross-validation method as implemented in the \texttt{mgcv} package. In keeping with the recommendation in \citet{Zhan:Huan:Lv:stat:2010}, the bandwidth for inferences in the GLRT approach was modified to be $\hat h_{\rm{opt}} \times n^{-1/20}$, where $\hat h_{\rm{opt}}$ was the estimated optimal bandwidth for model fitting. For each dataset, 200 bootstraps were used for the GLRT method due to its slow computation speeds (see average run times in Table \ref{tab:exp1}).

Table \ref{tab:exp1} displays the Type 1 error rates at nominal 1\%, 5\% and 10\% levels for simultaneously dropping $1,3,5$ and $7$ zero index coefficients using the GLRT, Wald test and the proposed PLRT approach from Proposition \ref{prop:LRT}. Here, dropping 1 covariate refers to testing $\beta_{10} = 0$, dropping 3 refers to testing $\beta_8 = \beta_9 = \beta_{10} = 0$, dropping 5 refers to testing $\beta _6 = \cdots = \beta_{10} = 0$ and dropping 7 refers to testing $\beta_4 = \cdots = \beta_{10} = 0$ simultaneously. Note that $\beta_3, \beta_4, \ldots, \beta_{10}$ are all exchangeable, so there is no loss of generality in defining hypotheses in this sequential manner.

From Table \ref{tab:exp1} we see that the proposed PLRT provides substantially less biased Type 1 error rates than those of the GLRT and Wald tests for both periodicities. While Type 1 error rates of both the Wald and PLRT methods approach nominal levels as the sample size increases, the PLRT always exhibits comparable, if not superior, performance throughout. Note that it was not feasible to run the GLRT on sample sizes of $n=400$ due to its extremely slow computation speed (see next paragraph). The Type 1 error rates in Table \ref{tab:exp1} suggest that the proposed PLRT can perform well for parameter inferences in Gaussian single-index models. 

Also displayed in Table \ref{tab:exp1} are the average computer run times for simultaneously dropping 7 covariates, $\beta_4 = \cdots = \beta_{10} = 0$, for each synthetic dataset using each of the three methods. We see that the computation times for the PLRT approach are comparable to that of the simple plug-in Wald test, but are over two orders of magnitude faster than the GLRT approach. Indeed, the long computation times for the GLRT make it rather infeasible for use in practice, taking over 83 minutes on average to analyze a {\it single} dataset of sample size $n=100$, and over 200 minutes to analyze a single dataset of sample size $n=400$. In contrast, the proposed PLRT approach does not require bootstrapping to approximate the null distribution of the test statistic, making it much more computationally efficient. This, coupled with its superior accuracy, makes it more appealing to use in practice.

\begin{table}
\tbl{Continuous responses with sinusoidal means - Type 1 error rates (\%) for simultaneously dropping 1, 3, 5 and 7 covariates, and average run times for simultaneously dropping 7 covariates, using GLRT, Wald and PLRT methods. Sample sizes $n = 100$ and $400$. $N = 1000$ simulations in each setting.}
{\begin{tabular}{lllrrrrrrrrrrrrr}\toprule
         &     &      & \multicolumn{12}{c}{\underline{Nominal significance levels (\%)}} \\
         &     &      & \multicolumn{3}{c}{\underline{Drop 1 covariate}} & \multicolumn{3}{c}{\underline{Drop 3 covariates}} & \multicolumn{3}{c}{\underline{Drop 5 covariates}} & \multicolumn{3}{c}{\underline{Drop 7 covariates}} & run time \\
a        & $n$ & method& 1   &  5  &   10  &  1  & 5   & 10    & 1   & 5   & 10    & 1   & 5    & 10   & (mins) \\
\colrule
$\pi/2$  & 100 & GLRT  & 7.0 & 13.2& 20.0  & 6.8 & 12.8& 16.6  & 5.6 & 8.2 & 12.4  & 8.2 & 11.2 & 15.0 & 83.10 \\
		 &	   & Wald  & 2.3 & 6.7 & 12.1  & 2.5 & 8.5 & 14.3  & 3.0 & 9.8 & 15.8  & 3.1 & 10.5 & 17.7 & 0.28 \\
         &     & PLRT  & 1.6 & 6.2 & 11.9  & 2.0 & 7.5 & 13.0  & 1.6 & 7.4 & 14.4  & 1.4 & 7.1  & 13.2 & 0.33 \\
         & 400 & GLRT & \multicolumn{12}{c}{not feasible} & $>200.00$ \\
         &    & Wald  & 1.7 & 6.6 & 10.4  & 2.0 & 7.4 & 14.8  & 2.9 & 8.9 & 14.9  & 3.0 & 9.5  & 15.6  & 0.48 \\
         &     & PLRT  & 1.4 & 5.9 & 9.7   & 1.2 & 6.1 & 12.2  & 1.8 & 7.2 & 13.4  & 1.2 & 6.7  & 11.9 & 0.55 \\
$3\pi/4$ & 100 & GLRT  & 5.2 & 10.0& 14.4  & 6.0 & 9.6 & 14.8  & 5.4 & 10.8& 16.2  & 10.4& 15.6 & 19.8 & 94.04 \\
		 &	   & Wald  & 1.9 & 6.2 & 12.2  & 2.8 & 7.6 & 12.4  & 3.1 & 8.7 & 14.1  & 3.5 & 9.9  & 16.0 & 0.66 \\
         &     & PLRT  & 2.3 & 6.2 & 12.6  & 1.9 & 6.9 & 11.0  & 1.9 & 6.5 & 13.1  & 1.7 & 7.6  & 12.6 & 0.71 \\
         & 400 & GLRT & \multicolumn{12}{c}{not feasible} & $>250.00$\\
         &        & Wald  & 1.1 & 5.8 & 11.6  & 1.3 & 7.3 & 13.2  & 1.6 & 7.5 & 13.6  & 1.5 & 8.0  & 14.1 & 0.87 \\
         &     & PLRT  & 1.1 & 5.6 & 10.8  & 0.9 & 6.4 & 11.9  & 0.9 & 6.1 & 11.1  & 0.9 & 5.5  & 10.6 & 0.98 \\
\botrule
\end{tabular}}
\label{tab:exp1}
\end{table}

We also looked at the accuracy of the equivalent standard error (\ref{eq:se}) obtained by inverting the PLRT. The simulation standard deviations, average Wald-based standard errors, and average equivalent standard errors of $\hat \beta_1$ and $\hat \beta_2$ for estimating the two non-zero coefficients $\beta_1$ and $\beta_2$ are given in the left half of Table \ref{tab:exp3}. These results suggest that the PLRT provides both accurate Type 1 errors for testing zero coefficients and accurate equivalent standard errors for inferences on non-zero coefficients.

\begin{table}
\tbl{Simulation ``true" standard errors ($\times 10^{-2}$), average Wald-based standard errors ($\times 10^{-2}$), and average equivalent standard errors ($\times 10^{-2}$) obtained by inverting the PLRT for $\hat \beta_1$ and $\hat \beta_2$ for simulated continuous data with sinusoidal means (Section 4.1) and simulated binary data with non-canonical means (Section 4.2). $N = 1000$ simulations in each setting.}
{\begin{tabular}{lllccclllcc}\toprule
\multicolumn{5}{c}{\underline{\hspace{8mm}Continuous responses\hspace{8mm}}} & & \multicolumn{5}{c}{\underline{\hspace{18mm}Binary responses\hspace{18mm}}} \\
\vspace{-2mm} \\
model        & $n$ & method & se$(\hat \beta_1)$ & se$(\hat \beta_2)$ && model     & $n$ & method & se$(\hat \beta_1)$ & se$(\hat \beta_2)$ \\
\colrule
$a=\pi/2$  & 100 & true   & 0.98	             & 1.92	              && c-log-log & 350 & true   & 3.43    & 6.91 \\
         &     & Wald   & 0.93	             & 1.84	              &&           &     & Wald   & 3.08    & 6.07 \\
         &     & PLRT   & 0.96	             & 1.90               &&           &     & PLRT   & 3.15    & 6.24 \\
         & 400 & true   & 0.43	             & 0.85	              &&           & 700 & true   & 2.30    & 4.53 \\
         &     & Wald   & 0.42	             & 0.84	              &&           &     & Wald   & 2.20    & 4.36 \\
         &     & PLRT   & 0.42	             & 0.87               &&           &     & PLRT    & 2.21    & 4.41 \\
$a=3\pi/4$ & 100 & true   & 0.62	             & 1.25	              && unimodal  & 350 & true   & 3.80    & 7.11 \\
         &     & Wald   & 0.71	             & 1.32               &&           &     & Wald   & 3.45    & 6.61 \\
         &     & PLRT   & 0.64	             & 1.29	              &&           &     & PLRT   & 3.43    & 6.72 \\
         & 400 & true   & 0.29	             & 0.58	              &&           & 700 & true   & 2.42    & 4.70 \\
         &     & Wald   & 0.28	             & 0.57	              &&           &     & Wald   & 2.43    & 4.75 \\
         &     & PLRT   & 0.30	             & 0.57               &&           &     & PLRT   & 2.38    & 4.71 \\
         &     &        &                    &                    && monotonic & 350 & true   & 4.33    & 9.17 \\
         &     &        &                    &                    &&           &     & Wald   & 3.96    & 8.14 \\
         &     &        &                    &                    &&           &     & PLRT   & 4.10    & 8.56 \\
         &     &        &                    &                    &&           & 700 & true   & 2.67    & 5.67 \\
         &     &        &                    &                    &&           &     & Wald   & 2.70    & 5.58 \\
         &     &        &                    &                    &&           &     & PLRT   & 2.74    & 5.74 \\
\botrule
\end{tabular}}
\label{tab:exp3}
\end{table}

\subsection{Binary responses with non-canonical mean curves}
We also compared the performance of the GLRT, Wald and PLRT approaches on binary data generated from the following three models:
\begin{enumerate}[leftmargin=*]
\setlength\itemsep{0.1em}
\item c-log-log: \hspace{13mm}
$P(Y_i = 1 \mid \bm{x}_i) = 1 - \exp(-\exp(\bm{x}_i^T \bm{\beta}))$ ;
\item
Unimodal:  \hspace{10.5mm}
$\mbox{logit}\left\{P(Y_i = 1 \mid \bm{x}_i) \right\}= -0.05(0.5 - 4 \bm{x}_i^T \bm{\beta})^2 + 0.8$ ;
\item  
Monotonic: \hspace{10mm}
$ \mbox{logit}\left\{P(Y_i = 1 \mid \bm{x}_i)\right\} = \exp(5 \bm{x}_i^T \bm{\beta} - 2)/ \{1+\exp(5 \bm{x}_i^T \bm{\beta} - 3)\} - 1.5 $;
\end{enumerate}

In each of the above settings, the sample size was set to either $n=350$ or $700$, corresponding to moderate and large sample sizes for binary data, respectively. The true index coefficients were set to $\bm{\beta} = (\beta_1,\beta_2,\beta_3,\beta_4)^T = (2,1,0,0)^T/\sqrt{5}$, and each covariate in $\bm{x}_i$ were simulated independently from a uniform distribution on $(-2,2)$. A total of $N=1000$ simulations were carried out for each setting.

For each simulated dataset, a binary GSIM model was fit using either local linear estimation for the GLRT approach, or penalized truncated cubic splines for the Wald and PLRT approaches. Again, the bandwidth for the local linear approach was chosen via the default least-squares cross-validation method as implemented in the \texttt{np} package, while the smoothing parameter for penalized cubic splines was chosen via the default cross-validation method as implemented in the \texttt{mgcv} package. In keeping with the recommendation in \citet{Zhan:Huan:Lv:stat:2010}, the bandwidth for inferences in the GLRT approach was again modified to be $\hat h_{\rm{opt}} \times n^{-1/20}$, where $\hat h_{\rm{opt}}$ was the estimated optimal bandwidth for model fitting. For each dataset, 200 bootstraps were again used for the GLRT method due to its slow computation speeds (see average run times in Table \ref{tab:exp234}).

\begin{table}
\tbl{Binary responses with non-canonical mean models - Type 1 error rates (\%)  and average run times (minutes) for simultaneously dropping 1 and 2 covariates using the GLRT, Wald and PLRT approaches. Sample sizes $n = 350$ and $700$. $N=1000$ simulations in each setting.}
{\begin{tabular}{lllrrrrrrr}\toprule
          &      &        & \multicolumn{6}{c}{\underline{Nominal significance levels (\%)}} \\
          &      &        & \multicolumn{3}{c}{\underline{Drop 1 covariate}} & \multicolumn{3}{c}{\underline{Drop 2 covariates}} & run time \\
model     & $n$  & method & 1   &  5  &   10   & 1   & 5     & 10   & (mins)         \\ 
\colrule
c-log-log & 350  & GLRT   & 1.8 & 10.0 & 21.0  & 2.2 & 11.6  & 22.0 & 105.93 \\
          &      & Wald   & 2.7 & 8.2  & 12.5  & 3.3 & 10.7  & 16.0 & 0.06   \\
          &      & PLRT   & 2.1 & 6.4  & 11.3  & 1.3 & 7.7   & 12.8 & 0.10   \\
          & 700  & GLRT   & 1.4 & 6.4  & 14.4  & 1.4 & 10.0  & 19.0 & 157.00 \\
          &      & Wald   & 2.0 & 5.8  & 10.6  & 1.7 & 6.7   & 11.9 & 0.08   \\
          &      & PLRT   & 1.2 & 4.9  & 9.7   & 1.3 & 5.6   & 10.9 & 0.13   \\
unimodal  & 350  & GLRT   & 7.8 & 17.4 & 21.8  & 7.8 & 15.4  & 21.2 & 75.02  \\
          &      & Wald   & 2.3 & 6.9  & 13.3  & 2.5 & 7.4   & 13.4 & 0.07   \\
          &      & PLRT   & 1.7 & 6.5  & 12.8  & 1.6 & 6.8   & 11.9 & 0.20   \\
          & 700  & GLRT   & 15.6& 18.2 & 20.2  & 12.8& 14.4  & 16.2 & 180.68 \\
          &      & Wald   & 1.1 & 5.8  & 10.4  & 1.2 & 4.9   & 10.7 & 0.10   \\
          &      & PLRT   & 1.0 & 5.6  & 10.3  & 1.1 & 4.8   & 9.9  & 0.24   \\
monotonic & 350  & GLRT   & 2.0 & 6.8  & 12.2  & 1.8 & 8.2   & 14.0 & 68.05 \\
          &      & Wald   & 2.5 & 8.1  & 15.1  & 3.8 & 10.1  & 16.0 & 0.08   \\
          &      & PLRT   & 1.7 & 6.7  & 12.5  & 2.6 & 7.3   & 13.2 & 0.19   \\
          & 700  & GLRT   & 5.6 & 7.6  & 12.8  & 4.6 & 9.6   & 13.4 & 154.60 \\
          &      & Wald   & 1.6 & 7.6  & 13.7  & 1.8 & 8.4   & 13.7 & 0.12   \\
          &      & PLRT   & 1.6 & 6.9  & 12.9  & 1.6 & 7.0   & 11.9 & 0.29   \\
\botrule
\end{tabular}}
\label{tab:exp234}
\end{table}

Table \ref{tab:exp234} displays the Type 1 error rates at nominal 1\%, 5\% and 10\% levels for simultaneously dropping 1 and 2 zero index coefficients using the GLRT, Wald test and the proposed PLRT approach from Proposition \ref{prop:LRT}. The results demonstrate that the proposed PLRT exhibits substantially less biased Type 1 error rates than those of the GLRT and Wald tests for all three mean models and for both sample sizes. The Wald test and PLRT both approach their nominal rates as the sample size increases, but the GLRT actually diverges. The particularly poor performance of the GLRT approach reflects the fact that it was designed for single-index models with additive errors and constant variance -- here, even bootstrapping the test statistic does not provide a good enough approximation to the null distribution when the data are binary. 

Also displayed in Table \ref{tab:exp234} are the average computation times for simultaneously dropping 2 covariates (i.e., $\beta_3 = \beta_4 = 0$) for each synthetic dataset using each of the three methods. We again see that the computation times for the PLRT approach are comparable to that of the simple plug-in Wald test, but over two orders of magnitude faster than the GLRT approach, which took over 68 minutes to analyze a single dataset of sample size $n=350$ and over 154 minutes to analyze a single dataset of sample size $n=700$. These computation times make the GLRT approach unusable in practice. In contrast, the proposed PLRT is both more accurate and computationally more efficient, making it more appealing to use in practice.

Finally, we also looked at the accuracy of the equivalent standard errors (\ref{eq:se}) for binary GSIMs obtained by inverting the PLRT. The simulation standard deviations, average Wald-based standard errors, and average equivalent standard errors of $\hat \beta_1$ and $\hat \beta_2$ for estimating the two non-zero coefficients $\beta_1$ and $\beta_2$ are given in the right half of Table \ref{tab:exp3}. These results again suggest that the PLRT provides both accurate Type 1 errors for testing zero coefficients and accurate equivalent standard errors for inferences on non-zero coefficients.

\section{Data analysis example}
\label{s4}
We apply the proposed PLRT method to make inferences on the relationship between the prevalence of bile duct hyperplasia in rats and 5 covariates, namely, gender, dose level, initial weight, cage position and age at death. The response is a binary variable, with $y = 1$ and $y = 0$ denoting the presence and absence of nonlethal lesions in the bile duct at death, respectively. The dataset consists of 319 samples and comes from  \citet{Dinse:1984}.

\citet[Section 6.4.1]{Green:1994} analyze the subset of male rats using a binary GSIM implemented via natural cubic splines with a fixed smoothing parameter. However, no standard errors or inferences for the index coefficients are provided. Here, we give a full analysis of the dataset by fitting a binary GSIM, with smoothing parameter chosen automatically by the \texttt{gam} function, computing standard errors and assessing the relative importance of each covariate. The R code for carrying out these calculations is provided in the Online Supplement.

Estimated index coefficients from the fitted model, along with standard errors and $p$-values based on both the PLRT and the usual plug-in estimator of variance, are displayed in Table \ref{tab:rat}. We see that inferences based on the two methods are qualitatively different here. For example, the PLRT suggests that dose level is a more important predictor than gender, but Wald-tests suggest the opposite. Age at death is not significant according to the Wald-test, but it is highly significant according to the PLRT. Because the PLRT exhibits substantially less bias in our simulations, we argue that they should be more reliable here.

\begin{figure}
\center
\includegraphics[scale = 0.6, trim={0 10 0 50}, clip]{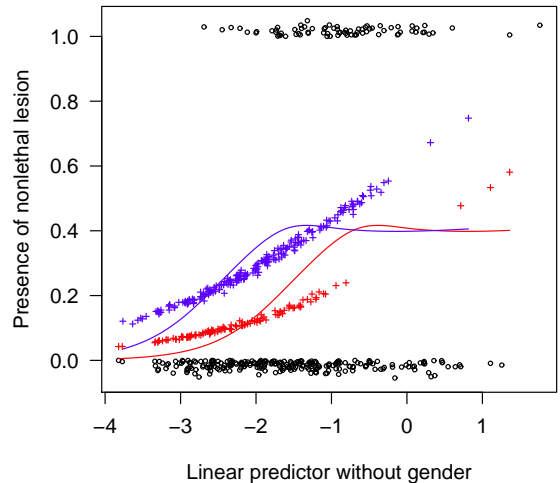}
\vspace{-3mm}
\caption{Scatterplot (with jitter) of tumour presence data, with fitted mean curves using generalized single-index (lines) and logistic regression (+ signs) models, for male (blue) and female (red) rats.}
\label{fig:rat}
\end{figure}

A logistic regression model was also fit to the data for comparison, with the corresponding estimates, standard errors and $p$-values displayed in Table \ref{tab:rat}. From Figure~\ref{fig:rat}, we find that the logistic model may be inadequate in capturing the functional relationship between the covariates and the tumour prevalence of rats. Specifically, the estimated mean curves obtained from the nonparametric GSIM suggest that the probability of tumour presence may increase up to some threshold but stays comparatively flat thereafter. This trend is not captured by the logistic model.

\begin{table}
\tbl{Rats tumour prevalence data analysis -- estimated coefficients, standard errors (se) and $p$-values based on profile likelihood ratio tests (PLRT) and Wald tests using plug-in estimator of variance.}
{\begin{tabular}{lrrrrrrrrr}\toprule
                     & \multicolumn{5}{c}{\underline{Generalized single-index model}}            & \multicolumn{3}{c}{\underline{Logistic regression}} \\
            & & \multicolumn{2}{c}{\underline{\hspace{3mm} PLRT\hspace{3mm}}} & \multicolumn{2}{c}{\underline{\hspace{3mm}Wald\hspace{3mm}}} \\ 
Covariate      & $\hat{\beta}$ & se & $p$ & se    & $p$      & $\hat{\beta}$ & se & $p$ \\ 
\colrule
Gender         & 0.945 & 0.458 & 0.040    & 0.054 & $<$0.001 & 1.127  & 0.431 & 0.009 \\ 
Dose level     & 0.258 & 0.090 & 0.005    & 0.126 & 0.042    & 0.152  & 0.082 & 0.061 \\ 
Initial weight & 0.002 & 0.006 & 0.707    & 0.016 & 0.880    & -0.003 & 0.009 & 0.753 \\ 
Cage position  & 0.200 & 0.099 & 0.044    & 0.129 & 0.121    & 0.131  & 0.097 & 0.177 \\ 
Age at death   & -0.034& 0.009 & $<$0.001 & 0.018 & 0.061    & -0.024 & 0.007 & 0.001 \\ 
\botrule
\end{tabular}}
\label{tab:rat}
\end{table} 

\section{Discussion}
The proposed PLRT approach is demonstrated to be both conceptually and computationally simple to implement, invariant to identifiability constraints, and can exhibit substantially less bias than standard Wald tests and the recently proposed GLRT method for inferences on the index parameters in GSIMs. Moreover, the computational times for the PLRT are comparable to the simple plug-in Wald test, and over two orders of magnitude faster than the GLRT. We believe that the accuracy of the PLRT can be further improved upon using Bartlett-type corrections. The method can also be extended to partially linear single-index models. These are topics for future research.

\vspace{-1mm}

\section*{Acknowledgements}
\vspace{-3mm}
We thank Bret Hanlon, Mark Hannay, the associate editor and two anonymous referees for comments and suggestions that improved the paper.

\vspace{-2mm}

\section*{Disclosure statement}
\vspace{-3mm}
No potential conflict of interest was reported by the authors.

\vspace{-2mm}

\section*{Supplemental material}
\label{SM}
\vspace{-3mm}
The Online supplement includes R code, another data analysis example and a proof of Proposition 1.

\vspace{-2mm}


\vspace{-5mm}

\appendices
\section*{Appendix}
\label{App}
\vspace{-3mm}
The results in Section \ref{s2} hold under the following regularity conditions:
\vspace{-2mm}
\begin{enumerate}[label=A\arabic*.]\setlength\itemsep{-0.1em}
  \item The parameter space $\Theta$ is compact.
  \item
The Fisher information matrix
\begin{eqnarray*}
I(\bm{\theta}) = E \left\{ \left[\frac{\partial \log f(y; \bm{\theta})}{\partial \bm{\theta}} \right] \left[\frac{\partial \log f(y; \bm{\theta})}{\partial \bm{\theta}} \right]^T \right\}
= E \left[b''(g(x; \bm{\theta})) \frac{\partial g(x; \bm{\theta})}{\partial \bm{\theta}} \frac{\partial g(x; \bm{\theta})}{\partial \bm{\theta}^T} \right]
\end{eqnarray*}
is finite and positive definite at $\bm{\theta} = \bm{\theta}_*$.
  \item For $\bm{\theta}$ in some neighbourhood of $\bm{\theta}_*$, there exist functions $M_{jkl}$ such that
$$
\Bigg| \frac{\partial^3 \log f(y; \bm{\theta}) }{\partial \bm{\theta}_j \partial \bm{\theta}_k \partial \bm{\theta}_l} \Bigg| \leq M_{jkl}(x,y),
$$
and $E_{\bm{\theta}_*}[M_{jkl}(x,y)] < \infty$ for all $j,k,l$.
\end{enumerate}

\end{document}